\documentclass[12pt]{article}
\usepackage[utf8]{inputenc}
\usepackage[english]{babel}
\usepackage{amsmath}
\usepackage{amsfonts}
\usepackage{amssymb}
\usepackage{cite}

\newcommand{\be}{\begin{equation}}
\newcommand{\ee}{\end{equation}}
\newcommand{\bea}{\begin{eqnarray}}
\newcommand{\eea}{\end{eqnarray}}

\newcommand{\lb}{\label}
\newcommand{\p}[1]{(\ref{#1})}

\begin{document}

\begin{titlepage}

\vspace*{0.7cm}

\begin{center}

{\LARGE\bf On the realization of infinite (continuous) }

\vspace{0.3cm}

{\LARGE\bf spin
field representations in AdS${}_{\mathbf{4}}$ space}

\vspace{2cm}

{\large\bf I.L.\,Buchbinder$^{1,2}$\!\!,\ \ \  S.A.\,Fedoruk$^1$\!\!,\ \ \
A.P.\,Isaev$^{1,3}$\!\!,\ \ \ M.A.\,Podoinitsyn$^{1}$}

\vspace{2cm}

\ $^1${\it Bogoliubov Laboratory of Theoretical Physics,
Joint Institute for Nuclear Research, \\
141980 Dubna, Moscow Region, Russia}, \\
{\tt buchbinder@theor.jinr.ru, fedoruk@theor.jinr.ru,
isaevap@theor.jinr.ru, mpod@theor.jinr.ru}

\vskip 0.5cm

\ $^2${\it Center of Theoretical Physics,
Tomsk State Pedagogical University, \\
634041 Tomsk, Russia}, \\

\vskip 0.5cm

\ $^3${\it Faculty of Physics, Lomonosov Moscow State University, \\
119991 Moscow, Russia}

\end{center}

\vspace{1.5cm}

\begin{abstract}

We study the symmetry properties of infinite spin fields in $\rm{AdS}_4$ space which are involved in the Lagrangian model
proposed in {\tt arXiv:2403.14446\,[hep-th]} where the main role is played by operator constraints. It is shown that the conditions defining infinite spin states in $\rm{AdS}_4$ space are $\mathrm{SO}(2,3)$-invariant. It is found that in the model under consideration the Casimir operators are completely fixed by the constraint operators and only one of the Casimir operators is independent. It is shown that in this model, infinite spin fields in $\rm{AdS}_4$ space are described by the most degenerate representations of the $\mathrm{SO}(2,3)$ group.

\end{abstract}

\vspace{2cm}



\vspace{1cm}

\end{titlepage}

\newpage

\setcounter{footnote}{0}

\setcounter{equation}{0}

\section{Introduction}

One of the relatively recent trends in higher spin field theory\footnote{Current state of the higher spin field problem is considered, e.g., in the reviews  \cite{reviewsV,Snowmass,Ponomarev}.} is the study of the Lagrangian formulation of infinite
(continuous) field models in $\rm{AdS}$ space. Various approaches to this formulation are discussed in refs.
\cite{Met,Met1,Zin,Met1-1,Met1-2,Met2,Met2-1,BFIK-24,BFIK-24_1,JO}.\footnote{The review of the infinite (continuous) spin field models is given in ref. \cite{BekSk}, for more recent references see e.g., \cite{BFIK-24_1}.}
In particular, a full classification of possible unitary and non-unitary continuous spin
field models was given in the paper by R.\,Metsaev \cite{Met2} within the light-cone formalism.

In a recent paper \cite{BFIK-24_1}, a completely consistent covariant Lagrange formulation for bosonic infinite spin field in $\rm{AdS}_4$ space was constructed in the framework of the BRST approach to higher spin fields. The formulation is realized in Fock space vectors $|{\Psi}\rangle$
where the BRST charge is constructed based on a set of constraint operators satisfying the closed algebra (see the operators \p{L-0}-\p{U} bellow).
The Lagrangian includes the real continuous parameter $\boldsymbol{\mu}$
and in the flat limit this Lagrangian is reduced to the known infinite spin field Lagrangian in Minkowski space \cite{BKT}
in which $\boldsymbol{\mu}$ defines the corresponding Poincar\'{e} representation.
The on-shell structure of the theory under consideration can also be formulated in terms of the field $\Phi$ which
depends on additional spinor variables and obeys all the same constraints as the formulation in the Fock space (see (\ref{IrRep-1} below) encoded in the BRST charge.
The relation between the descriptions in terms of Fock space vectors $|{\Psi}\rangle$ and on-shell fields $\Phi$
was discussed in detail in \cite{BKT} for infinite (continuous) spin representations in flat space.
Then natural questions arise as
to what representation of the symmetry group of $\rm{AdS}$ space such a formulation corresponds
and how it relates to Metsaev’s classification.
In this paper, we answer these questions.
First, we show that the field
that satisfies the system of constraints obtained within the Lagrangian formulation in the paper \cite{BFIK-24_1} corresponds
to the most degenerate continuous unitary representation of the Lie algebra $\mathfrak{so}(2,3)$.\footnote{The notion of the most degenerate representation is based on the Gelfand theorem \cite{Gelf} which states
that the number of independent invariant operators in the enveloping
algebra acting in the Hilbert space of functions on the domain $X$ can
be independent of the rank of the group. The most degenerate representation,
if it exists in given space, means that the number of these
operators is one. The construction of the most degenerate continuous
representations of the group  $\mathrm{SO}(p,q)$ is discussed in the papers
\cite{LNR-1,LNR-2,LN}, see also \cite{BR}.}
Second, we show that the representations of the $\mathrm{SO}(2,3)$ group realized on such a field correspond to the $\mathrm{SO}(2,3)$ representations in Metsaev's classification.
In this case, the eigenvalues of the  $\mathfrak{so}(2,3)$ Casimir operators
are determined by the dimensionless quantity $\boldsymbol{\mu}R$ where $R$ is the radius of $\rm{AdS}_4$ space.

The paper is organized as follows. In Section\,2, we briefly consider a reformulation in terms of the on-shell field $\Phi$  the results obtained in $AdS_4$ space \cite{BFIK-24_1}, similar to what was done  in flat space  \cite{BKT}.
Here the on-shell infinite spin field in $\rm{AdS}_4$
space is defined, the operator constraints $L_0,L,L^{+}$ are introduced, their algebra is written down and the flat limit is discussed.
In Section\,3, the geometric structure of $\rm{AdS}_4$ space is described in a form convenient for further consideration.
In Section\,4, we show that the operator constrains are invariants of the $\mathrm{SO}(2,3)$ group
and one more invariant operator $(\Pi)^{2}$ is introduced.
In Section\,5, we construct the Casimir operators $\mathcal{C}_2$ and $\mathcal{C}_4$ of the $\mathfrak{so}(2,3)$ algebra acting in the space of fields $\Phi$ under the conditions (\ref{IrRep-1}).
In doing so, we obtain two important statements. First, we prove that the Casimir operators $\mathcal{C}_2$ and $\mathcal{C}_4$
are completely expressed in terms of the operators of the constraints (\ref{IrRep-1}) under consideration.
Second, we show that only one of these two Casimir operators is independent. This exactly corresponds to the definition of
the most degenerate representation for the $\mathrm{SO}(2,3)$ group.
The summary is a collection of results.

\setcounter{equation}{0}

\section{Infinite spin field in $\rm{AdS_4}$ space}

We begin with a brief summary of the basic aspects related to the Lagrangian formulation of the
infinite spin field in $\rm{AdS_4}$ space .

A consistent Lagrangian formulation for an infinite spin field in flat space was constructed
in paper \cite{BKT}. Following this paper, generalization for the infinite spin theory in $\rm{AdS_4}$ space  is conveniently realized in terms of fields
of the form
\be
\lb{Psi}
{\Phi}={\Phi}(x,c,\bar c)\,,
\ee
where
$x^\mu$, $\mu=0,1,2,3$ are  the coordinates  on $\rm{AdS_4}$ space and
$c^\alpha$, $\bar{c}_{\dot{\alpha}}=\epsilon_{\dot{\alpha}\dot{\beta}}\bar{c}^{\dot{\beta}}$, $\bar{c}^{\dot{\alpha}}=(c^\alpha)^\dagger$
are the components of the commuting Weyl spinor.\footnote{The
two-component Weyl spinor indices are raised and lowered by
$\epsilon_{\alpha\beta}=-\epsilon^{\alpha\beta}$,
$\epsilon_{\dot\alpha\dot\beta}=-\epsilon^{\dot\alpha\dot\beta}$:
$\varphi^{\alpha}=\epsilon^{\alpha\beta}\varphi_{\beta}$,
$\bar\varphi^{\dot\alpha}=\epsilon^{\dot\alpha\dot\beta}\bar\varphi_{\dot\beta}$.
The relativistic Pauli matrices $\sigma^m_{\alpha\dot\beta}$ and
$\tilde\sigma_m^{\dot\alpha\beta}$ satisfy the basic relation $\sigma_m
\tilde{\sigma}_n + \sigma_n \tilde{\sigma}_m ={} -2 \eta_{mn}$ where
$\eta_{mn}=\mathrm{diag}(-1,+1,+1,+1)$. We use the matrices
$\sigma_{m n} = -\left(\sigma_{m} \tilde{\sigma}_n - \sigma_{n}
\tilde{\sigma}_m\right)/4$, $\tilde {\sigma}_{m n} =
-\left(\tilde{\sigma}_m \sigma_{n}  - \tilde{\sigma}_n \sigma_{m}
\right)/4$. Here and further the Latin indices take the values 0,1,2,3.}
The construction of fields (\ref{Psi}) is discussed in details in \cite{BKT}.
Also we use the operators
\be
\lb{vac}
a_\alpha=\frac{\partial}{\partial c^\alpha}\,, \qquad
\bar{a}^{\dot\alpha}=\frac{\partial}{\partial \bar{c}_{\dot{\alpha}}} \,,
\ee
satisfying the following commutation relations:
\be \lb{sn-1}
[a_{\alpha}, c^{\beta}] =
\delta_{\alpha}^{\beta} \,, \qquad
[\bar{a}^{\dot{\alpha}},\bar{c}_{\dot{\beta}}] =
\delta_{\dot{\beta}}^{\dot{\alpha}}\,.
\ee

The infinite spin states in $\rm{AdS_4}$ are described by
the fields (\ref{Psi}) subject to  the conditions
\be \lb{IrRep-1}
L_0 {\Phi} = 0 \,, \qquad L {\Phi} = 0 \,, \qquad
L^+ {\Phi} = 0 \,, \qquad U {\Phi} = 0 \, ,
\ee
with the constraint operators found in \cite{BFIK-24_1} in the form
\bea \lb{L-0}
L_0 &=& (D)^2 - \frac{1}{R^2}
\left ( N \bar{N} + N + \bar{N}\right)
+  \frac{2\boldsymbol{\mu}}{R} + \frac{1}{2 R^2} \, K^2 \,, \\[7pt]
\lb{L-n}
L &=& i \, (a\sigma^m\bar a) D_m - \boldsymbol{\mu} - \frac{1}{4 R}\, K^2 \,,  \\[7pt]
\lb{L-p}
L^+ &=& i \, (c\sigma^m\bar c) D_m - \boldsymbol{\mu} - \frac{1}{4 R}\, K^2 \,, \\[7pt]
\lb{U}
U &=& N- \bar N \,,
\eea
where the quantities $N$, $\bar{N}$, $K$ are defined by the expressions
\be \lb{sn-2}
N = c^{\alpha} a_{\alpha} \,, \qquad \bar{N} = \bar{c}_{\dot{\alpha}} \bar{a}^{\dot{\alpha}}
 \,, \qquad K= N + \bar{N}+2\,,
\ee
the  real parameter $\boldsymbol{\mu}$ is associated with the
infinite spin representation, the positive constant $R$ is the
curvature radius of $AdS_4$ space, \footnote{In \cite{BFIK-24_1},
the parameter $\kappa$ was used, which is equal to $\kappa =
-1/R^2$.} and
\be \lb{s-f-5}
D_m = e_m^\mu D_\mu\,,\qquad
D_\mu=\partial_\mu + \frac12\, \omega_{\mu}{}^{ k l}
\mathcal{M}_{kl} \ee
is the covariant derivative. Here $e_m^\mu$ is the
matrix inverse to the vierbein $e^m_\mu$, $e_m^\mu e^n_\mu=
\delta_n^m$, $e^m_\mu e_{m\nu}=g_{\mu\nu}$, the real field
$\omega_{\mu}{}^{ k l}(x)$ is the spin-connection and the operators
\be
\lb{sn-3}
\mathcal{M}_{m n} = M_{m n} + \bar{M}_{m n} \,, \qquad
M_{m n} := c^{\alpha}\, (\sigma_{m n})^{\;\; \beta}_{\alpha} \,
a_{\beta} \,, \quad \bar{M}_{m n} := \bar{c}_{\dot{\alpha}} \,
(\tilde{\sigma}_{m n})^{\dot{\alpha}}_{\;\;\dot{\beta}} \,
\bar{a}^{\dot{\beta}} \,,
\ee
are the anti-Hermitian generators,
$(\mathcal{M}_{m n})^\dagger={}-\mathcal{M}_{m n}$,  of the Lorentz
algebra $\mathfrak{so}(1,3)$. In \p{L-0}, the operator $(D)^2$ has
the form
\be \lb{d-sq}
(D)^2=D^m D_m + e^\mu_m\omega_{\mu}^{\;\;m l}
D_l = \frac{1}{\sqrt{-g}}\, D_\mu \sqrt{-g}g^{\mu\nu}D_\nu\,,
\ee
where $g=\det g_{\mu\nu}$. Relation \p{d-sq} allows one to call the
operator $(D)^2$ the generalized Laplace-Beltrami operator in
 space of vectors (\ref{Psi}). Additionally, in \p{L-n} and
\p{L-p}, we use the non-indexed notation: $(c\sigma^m\bar c):=
c^{\alpha} \sigma^m_{\alpha \dot{\alpha}} c^{\dot{\alpha}}$,
$(a\sigma^m\bar a) := a^{\alpha} \sigma^m_{\alpha \dot{\alpha}}
a^{\dot{\alpha}}$.

The operators \p{L-0}-\p{U} form a closed algebra in terms of commutators:
the only nonzero commutator has the form
\be
\lb{al2}
[L^+,L]
=
KL_0
+\frac{1}{R}\,(K+1)L
+\frac{1}{R}\,(K-1)L^+ \,.
\ee

Before going further, let us make a number of important comments  about the defining conditions
\p{IrRep-1}.

First, the operator $U$ in the last condition \p{IrRep-1} is the $\mathrm{U}(1)$ symmetry generator \cite{BFIK-24_1}.
Second, the limit $R\to \infty$ (together with $\omega_{\mu}{}^{ k
l}= 0$ and $e_m^\mu=\delta_m^\mu$) is well defined: it simply
switches off the external gravitational field. In this limit,
equations \p{IrRep-1} yield the conditions that define
the well-known representation of the Poinc\'are group with infinite spin in
flat space. Indeed, in the flat limit the operators \p{L-0}-\p{L-p}
take the form
\be \lb{flat}
\left. L_0 \right|_{R\to\infty} =
\partial^m\partial_m \,, \qquad \left. L \right|_{R\to\infty}= i \,
(a\sigma^m\bar a)\, \partial_m -\boldsymbol{\mu}\,,  \qquad \left.
L^+ \right|_{R\to\infty}= i \, (c\sigma^m\bar c)\, \partial_m
-\boldsymbol{\mu}\,.
\ee
In this case, the anti Hermitian generators
of the translations are $P_m=\partial_m$ and the square of the
Pauli-Lubanski pseudovector $ \displaystyle
W_m=-\frac12\,\epsilon_{mnkl}P^n \mathcal{M}^{kl} $ is equal to
\be\label{PL-val}
W^mW_m \ = {}-(c\sigma^m\bar c)(a\sigma^n\bar
a)\,\partial_m\partial_n \ +\ \left[\frac{N}{2}\left(\frac{N}{2}+1
\right)+ \frac{\bar N}{2}\left(\frac{\bar N}{2}+1 \right)
+\frac{N\bar N}{2}\right]\partial^m\partial_m\,.
\ee
Taking into account relations \p{flat}, one gets that the condition
$W^mW_m=\boldsymbol{\mu}^2$ holds for functions obeying the first
three equations  \p{IrRep-1} in the flat case $R\to\infty$. The last
condition in \p{U} removes the degeneracy arising from the use of
additional spinor variables in our approach.

Thus, in the flat case $R\to\infty$, the conditions  \p{IrRep-1} define the infinite spin field.
\footnote{Note that our formulation of infinite spin fields uses spinor coordinates as additional variables unlike the vector additional coordunates in the seminal papers \cite{Wig,BargWig} (see also \cite{BekSk} for review). Therefore, although our description of the field differs from the Bargman-Wigner formulation,
the correct value of the squared Pauli-Lubanski vector \p{PL-val} guarantees 
 a correct description of massless irreducible representations of infinite (continuous) spin in flat space (see discussion after \p{PL-val}). Besides, it was shown in \cite{BFIR} that the Bargman-Wigner formulation with vector additional coordinates is reproduced within the formulation with spinor additional coordinates if to carry out a spinor integral transform. We believe that the description using spinor additional coordinates is technically simpler.}
On the other hand, equations  \p{IrRep-1} for a nonzero curvature $\kappa = -1/R^2 \neq 0$ provide
a consistent dynamical description of such a system in $AdS_4$ space.

Our main aim further is to clarify which representations of the $\rm{AdS_4}$ symmetry group
correspond to the dynamical infinite spin field system constructed above.

\setcounter{equation}{0}

\section{Geometry of $\rm{AdS_4}$ space}

One of the ways to define the anti-de Sitter space $\rm{AdS_4}$ of the radius $R$ is to identify it
with the hypersurface
\be \lb{H-AdS}
X^{M}  X^{N} \eta_{MN} = - R^2
\ee
in five-dimensional two-time space $\mathbb{R}^{2,3}$ with the metric
$\eta_{MN} = \mathrm{diag} (-,+,+,+,-)$ and the coordinates $X^M$, $M=0,1,2,3,4$.
The manifold defined by the condition \p{H-AdS} is the coset $\mathrm{SO}(2,3)/\mathrm{SO}(1,3)$.
The symmetry generators of $\rm{AdS_4}$ space are \footnote{
Throughout this paper we use the
anti-Hermitean form of generators: for example, $(J_{MN})^\dagger ={} -J_{MN}$.
}
\be \lb{g-ads-g}
J_{MN} = X_M \frac{\partial}{\partial X^N}-X_N \frac{\partial}{\partial X^M}\,,
\ee
where $X_M=\eta_{MN}X^N$ form the $\mathfrak{so}(2,3)$ algebra:
\be \lb{com-rel}
[J_{MN},J_{KL}] =
\eta_{NK} J_{ML}+\eta_{ML} J_{NK}-\eta_{MK} J_{NL} - \eta_{NL} J_{MK}\,.
\ee
For the generators $J_{mn}$, $m=0,1,2,3$ together with the generators
\be \lb{n-gen}
P_{m} : = R^{-1} J_{m4} \,,
\ee
the algebra \eqref{g-ads-g} takes the form of the $\rm{AdS_4}$ symmetry algebra:
\be \lb{adsal}
\begin{array}{c}
[P_{m}, P_{n}] = R^{-2} J_{mn}  \,,   \\[7pt]
[J_{mn},P_{l}] = \eta_{n l} P_{m} - \eta_{m l} P_{n}  \, ,\qquad
[J_{mn}, J_{k l}] = \eta_{n k} J_{m l}  + \eta_{m l } J_{n k} - \eta_{n l} J_{m k} - \eta_{m k } J_{n l} \, ,
\end{array}
\ee
where $\eta_{mn}= \mathrm{diag}(-,+,+,+)$
and the operators $J_{mn}$ form the $\mathfrak{so}(1,3)$ subalgebra.

Another way to describe this space is to use four independent coordinates
$x^\mu$, $\mu=0,1,2,3$. Expressions for the coordinates $X^M$
in terms of the curved space coordinates $x^\mu$, resolving the condition \p{H-AdS}, can be chosen in the form
(see  e.g., \cite{BK})
\be\lb{Xx}
X^m=e_\mu^m(x)\, x^\mu \,,\qquad X^4=R\left(\frac{2}{G(x)}-1\right)\,,
\ee
where
\be\lb{Gx}
G(x)=1-\frac{\eta_{\mu\nu}x^\mu x^\nu}{4R^2}
\ee
and
\be\lb{ex}
e_\mu^m(x)=G^{-1}(x)\,\delta_\mu^m
\ee
is the vierbein.
The inverse matrix is formed by the vectors
\be\lb{e-x}
e^\mu_m(x)=G(x)\,\delta^\mu_m\,,
\ee
whereas the metric tensors are
\be\lb{gx}
g_{\mu\nu}(x)=G^{-2}(x)\,\eta_{\mu\nu}\,,\qquad g^{\mu\nu}(x)=G^{\,2}(x)\,\eta^{\mu\nu}\,.
\ee
Let us emphasize that in \p{Gx} and \p{gx} the tensors $\eta_{\mu\nu}= \eta^{\mu\nu}=
\mathrm{diag}(-,+,+,+)$ are constant despite the fact that they have the curved space indices.

In the case where the $\rm{AdS_4}$ metric quantities have the form \p{Gx}-\p{gx},
the spin connection is written as follows:
\be\lb{om-ads}
\omega_{m}{}^{kl}(x)=e^\mu_m \omega_{\mu}{}^{kl}(x)= R^{-2}\,{\delta_m^{[k}\delta_\nu^{l]}x^\nu} \,.
\ee
At the same time, we obtain the standard form of the $\rm{AdS_4}$  curvature tensor:
\be\lb{R-ads}
\mathcal{R}_{mn}{}^{kl}=e^\mu_m e^\nu_n \mathcal{R}_{\mu\nu}{}^{kl}(x)={}- R^{-2}\left(\delta_m^{k}\delta_n^{l}- \delta_m^{l}\delta_n^{k}\right)\,.
\ee

In the curved variable $x^\mu$ in  $\rm{AdS_4}$ space \p{H-AdS} the generators $J_{mn}$ defined in \p{g-ads-g}
are represented only by orbital generators $J_{mn}=\mathcal{L}_{mn}$ which have the form
(here $\eta_{m\mu}= \mathrm{diag}(-,+,+,+)$):
\be\lb{L-orb}
\mathcal{L}_{mn}=\left(\eta_{m\mu}\delta_n^\nu-\eta_{n\mu}\delta_m^\nu\right)x^\mu\,\frac{\partial}{\partial x^\nu}\,.
\ee
In this case, the generators \p{n-gen} take the form
\be \lb{s-t-1}
P_{m} = \left(1+\frac{\eta_{\nu\lambda}x^\nu x^\lambda}{4 R^2}\right) \delta^\mu_m\partial_\mu
- \frac{1}{2 R^2}\,\eta_{m\mu}x^\mu x^\lambda\partial_\lambda \,.
\ee
This expression \p{s-t-1} is represented as
\be \lb{s-f-0}
P_m = e^\mu_m  \left(\partial_\mu - \frac12\, \omega_\mu{}^{k l} \mathcal{L}_{kl} \right) ,
\ee
where $e^\mu_m$, $\omega_\mu{}^{k l}$ and $\mathcal{L}_{kl}$ are defined in \p{e-x}, \p{om-ads} and \p{L-orb}, respectively.

The operators \p{s-f-0} and \p{L-orb} realize the representation of the $\rm{AdS_4}$ algebra \p{adsal}
in the space of scalar functions depending on the curved coordinate $x^\mu$.
However, the infinite spin system under consideration is defined by conditions \p{IrRep-1} on
the Fock space vector \p{sn-1}, \p{vac}, where
the creation $c^\alpha$, $\bar{c}_{\dot\alpha}$ and annihilation $a_\alpha$, $\bar{a}^{\dot\alpha}$
operators act, carrying indices of the tangent space structural group $\mathrm{SO}(1,3)$.
This additional structure is taken into account by the replacement $\mathcal{L}_{mn}\ \to \ \mathcal{L}_{mn}+\mathcal{M}_{m n}$ in the operators  \p{s-f-0} and \p{L-orb}
where  $\mathcal{M}_{m n}$ are defined in \p{sn-3}.

Thus, as generators of the  $\mathrm{SO}(1,3)$ symmetry of the system under study, we consider the
operators
\be \lb{s-t-5}
J_{mn}=\mathcal{L}_{mn} + \mathcal{M}_{mn} \,,
\ee
where $\mathcal{L}_{mn}$ and $\mathcal{M}_{mn}$  are defined in \p{L-orb} and  \p{sn-3}, respectively.
Four residual generators are
\be \lb{s-f-1}
P_m = e^\mu_m  \left(\partial_\mu - \frac12\, \omega_\mu{}^{k l} J_{kl} \right) ,
\ee
where $J_{kl}$ are defined in \p{s-t-5}.
It is easy to check by direct calculations that the operators \eqref{s-t-5} and \eqref{s-f-1} obey relations \eqref{adsal} and thus form the $\mathfrak{so}(2,3)$ algebra.

Note that some of the commutators in \eqref{adsal} can be represented as
\be \lb{s-f-2}
[P_n, P_m] = {} - \frac12\, \mathcal{R}_{n m}{}^{k l} J_{k l}\,,
\ee
where $\mathcal{R}_{n m}{}^{k l}$ are defined in \p{R-ads}.

\setcounter{equation}{0}

\section{$\mathrm{SO}(2,3)$ symmetry of infinite spin field}

Having found expressions for the $\mathfrak{so}(2,3)$ symmetry generators,
we now analyze the $\mathrm{SO}(2,3)$ symmetry of the conditions \p{IrRep-1}
that determine the infinite spin field in $\rm{AdS_4}$ space.
The operator constraints in relations \p{L-0}-\p{U} are expressed in terms of the operator
\p{s-f-5}, which is written as follows:
\be \lb{s-f-5a}
D_m = e_m + \frac12 \, \omega_{m}{}^{ k l} \mathcal{M}_{kl} \,,
\ee
where $\omega_{m}{}^{ k l}=e^\mu_{m}\omega_{\mu}{}^{ k l}$ and the first term equals
\be \lb{fre}
e_{m} := e_{m}^{\mu} \partial_{\mu} \,.
\ee
The operators \p{fre} form the algebra
\be \lb{AnhC}
[e_m, e_n] = \mathcal{E}_{mn}{}^{ l} e_l \,,
\ee
where
\be \lb{frf}
\mathcal{E}_{mn}{}^{l} := (e_m e_{n}^{\mu} - e_n e_m^{\mu})\, e_{\mu}^l
\ee
is the anholonomy tensor, which is equal to
$\mathcal{E}_{mn}{}^{l}  = - x^\mu \eta_{\mu[m} \delta_{n]}^{l} /R^2$ in $\rm{AdS_4}$ space \p{ex}, \p{e-x}.
In these terms, the commutators of the operators \p{s-f-5a}
are written as follows:
\be \lb{s-f-6D}
[D_m, D_n] = \mathcal{E}_{mn}{}^{l} D_l +
\frac12\, \mathcal{R}_{mn}{}^{k l} \mathcal{M}_{k l}\,,
\ee
whereas the commutation relations of the operators \p{s-f-5a} with
the generators \eqref{s-t-5} and \eqref{s-f-1} of the $\rm{AdS_4}$ algebra $\mathfrak{so}(2,3)$
are
\be\lb{s-f-11}
[P_m, D_k] = \omega_{mk}{}^l D_l \,, \qquad
[J_{mn}, D_k] = \eta_{nk} D_{m} - \eta_{mk} D_n \,.
\ee
In obtaining relations \p{s-f-11}, the commutators
\bea \lb{s-f-3}
[\mathcal{L}_{mn}, e_{k}] &=& \eta_{nk} e_m - \eta_{m k} e_n \,, \\[7pt]
\lb{s-f-4}
[\mathcal{L}_{mn}, \omega_{klr}] &=& \eta_{nk}\omega_{mlr} - \eta_{mk}\omega_{nlr}
+\eta_{nl}\omega_{kmr}-\eta_{ml}\omega_{knr} +\eta_{nr}\omega_{kl m}-\eta_{mr}\omega_{kln} \,.
\eea
are used, as well as the equalities
\be \lb{s-f-9}
e_{(m} \omega_{n)}{}^{kl} = 0\,,\qquad
\omega_{(m}{}^{r[k} \omega_{n)r}{}^{l]} = 0 \,,
\ee
that are valid for  $\rm{AdS_4}$ space defined by the quantities \p{Gx}-\p{om-ads}.

Now, relations \p{s-f-11} together with the commutators
\be \lb{sn-6}
[\mathcal{M}_{m n}, (c\sigma_l\bar c)] = \eta_{n l} (c\sigma_m\bar c) - \eta_{m l} (c\sigma_n\bar c) \,, \qquad
[\mathcal{M}_{m n}, (a\sigma_l\bar a)] = \eta_{n l} (a\sigma_m\bar a) - \eta_{m l} (a\sigma_n\bar a)
\ee
and $[P_m, K] = [J_{mn}, K] =0$  lead to the statement that the operators $L$ and $L^+$ defined in \p{L-n} and \p{L-p} are
$\mathrm{SO}(2,3)$-invariant:
\be \lb{soLLP}
[P_n, L] = [P_n, L^+] = 0 \,, \qquad [J_{m n}, L] = [J_{m n}, L^+] = 0 \, .
\ee

In addition, using the commutation relations \p{s-f-11} and equalities \p{s-f-9} for the $\rm{AdS_4}$ metric and spin connection,
we obtain that the operator  $(D)^2$ defined in \p{d-sq} commutes with all generators of the $\rm{AdS_4}$ algebra $\mathfrak{so}(2,3)$:
\be \lb{pr-soLZ-1}
[P_m, (D)^2] = 0 \,, \qquad [J_{m n}, (D)^2] = 0 \,.
\ee
Note that the simplest way to prove the second relation in \eqref{pr-soLZ-1} is to use the following expression
for the operator  $(D)^2$:
\be
(D)^2 = D_m D^m + [P_m, D^m] \,.
\ee
Thus, the conditions \p{pr-soLZ-1} lead to the statement that the operator $L_0$ defined in  \p{L-0} is $\mathrm{SO}(2,3)$-invariant:
\be \lb{soLZ}
[P_m, L_0]  = 0 \,, \quad [J_{m n}, L_0] =  0 \, .
\ee
In addition, it is easy to show that
the operator \p{U} also commutes with the $\mathfrak{so}(2,3)$ algebra generators:
\be \lb{soLU}
[P_m, U]  = 0 \,, \quad [J_{m n}, U] =  0 \, .
\ee

Thus, the zero commutators \p{soLLP}, \p{soLZ}, \p{soLU} of all operators
in the set \p{L-0}-\p{soLU}
and all $\mathfrak{so}(2,3)$ generators \p{s-t-5}, \p{s-f-1} lead to a very strong conclusion:
the conditions \p{IrRep-1} defining the infinite spin field in $AdS_4$ space are
$\mathrm{SO}(2,3)$-invariant. As a result, we immediately face the question which representations of
the $\mathrm{SO}(2,3)$ group correspond to the Fock space vectors satisfying the conditions \p{IrRep-1}.
The next section is devoted to the discussion of this issue.

At the end of this section we introduce one more invariant operator of the algebra
$\mathfrak{so}(2,3)$, which will be useful for subsequent consideration.

First, using the equality
\be \lb{sn-12}
\mathcal{M}^{m n} \mathcal{M}_{m n} = M^{mn} M_{mn} + \bar{M}^{mn} \bar{M}_{mn} =  - N(N+2) - \bar{N}(\bar{N} + 2)\,,
\ee
we  can represent the operator $L_0$ \p{L-0} in the form:
\be \lb{LZinV}
L_0 = (D)^2 - \frac{1}{2 R^2}\, \mathcal{M}_{m n} \mathcal{M}^{mn} + \frac{2\left(1+\boldsymbol{\mu} R \right)}{R^2} \,.
\ee

Second,  let us define the vector operator
\be \lb{Pi}
\Pi_m := \mathcal{M}_{m n} D^n \,,
\ee
which has the following commutation relations with the $\mathfrak{so}(2,3)$ generators:
\be \lb{Pi-cr}
[P_m, \Pi_k] = \omega_{mk l} \Pi^l \,, \qquad [J_{m n}, \Pi_k] = \eta_{n k} \Pi_m - \eta_{m k} \Pi_n \, .
\ee
Using these commutators and equalities \p{s-f-9} for the $\rm{AdS_4}$ spin connection,
we obtain that the operator
\be \lb{PIsq}
(\Pi)^2 := \Pi_m \Pi^m - \mathcal{M}_{m n} \omega^{m n l} \Pi_l
\ee
commutes with all $\mathfrak{so}(2,3)$ generators:
\be\lb{PIsq-PJ}
[P_m, (\Pi)^{2}] = 0\,, \qquad [J_{m n}, (\Pi)^2] = 0  \,
\ee
and hence it is an invariant of the $\mathfrak{so}(2,3)$ algebra.

\section{Casimir operators and irreducible unitary representations}

The  $\mathfrak{so}(2,3)$ algebra is characterized by two Casimir operators. In this section we show
that in Fock space of the vectors satisfying the conditions \p{IrRep-1}
the Casimir operators are expressed through the operator
constraints \p{L-0}-\p{U} and describe the corresponding $\mathrm{SO}(2,3)$ irreducible representation.

\subsection{Second order Casimir operator}

The $\mathfrak{so}(2,3)$ algebra \p{com-rel} (or the same \p{adsal}) has the following
second order Casimir operator
\be \lb{cas}
\mathcal{C}_2 = {}-\frac{1}{2}\, J^{MN} J_{MN} = R^2 \, P^{m} P_{m} -\frac{1}{2}\, J^{m n} J_{m n}\,.
\ee
Direct calculations lead to the following explicit expression for this operator
\be \lb{DM-PJ}
P_m P^m - \frac{1}{2R^2}\, J^{m n} J_{m n} =
(D)^2 - \frac{1}{2R^2}\, \mathcal{M}^{m n} \mathcal{M}_{m n} \,.
\ee
Using this relation together with expression \p{LZinV}, one gets for the Casimir operator \p{cas}
\be \lb{L0-C2}
\mathcal{C}_2 = R^2{L}_0 -2\left(1+\boldsymbol{\mu} R \right)\,.
\ee
Thus, the Casimir operator $\mathcal{C}_2$ is defined by the operator
constraint $L_0$. Therefore, on the vectors \p{Psi} corresponding to the infinite spin field
it acts as the operator of multiplication by a constant $-2(1+\boldsymbol{\mu} R)$.

\subsection{Fourth order Casimir operator}

The analysis of the fourth-order Casimir operator is not so simple as in the case of the second order Casimir
operator.

The fourth order Casimir operator is defined as the square of the 5-vector
\be \lb{B1}
W_M := \frac{1}{8}\, \varepsilon_{M N K L I } J^{N K} J^{L I}\,.
\ee
After some transformations, one gets (see e.g., \cite{Met2})
\be \lb{B4}
\mathcal{C}_4 = W_{M} W^{M}  = \frac12 \, \mathcal{C}_2 \left(\mathcal{C}_2 + 3\right)
- \frac14\, J_{MN} J^{NK} J_{KL} J^{LM}\,,
\ee
where $\mathcal{C}_2$ is defined in  \eqref{cas}.

Let us now rewrite the last term in \p{B4} using the generators $J_{m n}$ and $P_m$ defined in \p{n-gen}.
We get
\bea \lb{B6}
\nonumber
J_{MN} J^{NK} J_{KL} J^{LM} &=&{}
- 4 R^2 \, P_m J^{m n} P^{k} J_{k n} + 2 R^4 \, (P_m P^m)^2 + 6R^2 \, P_n P^n
\\[7pt]
\lb{B5-1}
&&{}  - 4 J^{m n} J_{m n}
+ J_{mn} J^{nk} J_{kl} J^{lm} \,.
\eea
Relation \p{B5-1} is transformed as follows. First, one expresses $P_m P^m$ through
$\mathcal{C}_2$ using \p{cas}, then substitutes the result into \p{B4} and uses all above relations in
\p{B5-1}. Then one gets that
the first term in \p{B5-1} $\frac12 \, \mathcal{C}_2 \left(\mathcal{C}_2 + 3\right)$ is completely
canceled out. As a result, we obtain
\be \lb{casPJ}
 \mathcal{C}_4 = {}-\frac12 \,J^{mn} J_{mn}\, \mathcal{C}_2 + R^2  P_{m} J^{m n} P^{k} J_{k n}
 + \frac14 \,J^{mn} J_{mn} -  \frac18 \,(J^{mn} J_{mn})^2 -  \frac14\, J_{mn} J^{nk} J_{k l} J^{l m} \, .
\ee
This operator can be represented in the form (see also a similar expression in \cite{Met2}):
\be \lb{C4PI}
\mathcal{C}_4 = {}-\frac12\, \mathcal{M}^{mn} \mathcal{M}_{mn}\, \mathcal{C}_2 +
R^2  (\Pi)^2  - A(\mathcal{M})
\,,
\ee
where
\be \lb{A-C4PI}
A(\mathcal{M}):=
\frac54\, \mathcal{M}^{mn} \mathcal{M}_{mn}  + \frac18\, (\mathcal{M}^{mn} \mathcal{M}_{mn})^2 +
\frac14\, \mathcal{M}_{mn} \mathcal{M}^{nk} \mathcal{M}_{kl} \mathcal{M}^{lm} \,,
\ee
and the $\mathfrak{so}(2,3)$-invariant operator $(\Pi)^2$ is given in \eqref{PIsq}.

Let us distinguish in the fourth-order Casimir operator, the operators \p{L-0}-\p{U}.
To do this, we represent the operator $(\Pi)^2$ \eqref{PIsq} in the following form:
\be \lb{ap-D-1}
(\Pi)^2 = \mathcal{M}^{l m} \mathcal{M}_{l}{}^{n} \left(D_m D_n +
\omega_{m n k} D^k\right)\,.
\ee
After using equalities
\p{sn-12},
\be \lb{11}
\mathcal{M}_{m n} \mathcal{M}^{n l} \mathcal{M}_{l r} \mathcal{M}^{r m}=
-\mathcal{M}^{m n} \mathcal{M}_{m n}+
\frac14\,(\mathcal{M}^{m n} \mathcal{M}_{m n})^2 +N(N+2)\bar{N}(\bar{N}+2)\,,
\ee
\be \lb{ap-MM}
\mathcal{M}^{l (m} \mathcal{M}_{l}{}^{n)}=
{}S(N,\bar{N})\,\eta^{mn}
-(c\sigma^{(m}\bar c)(a\sigma^{n)}\bar a)\,,
\ee
where
\be \lb{S-ap-MM}
S(N,\bar{N}):=
{}-\frac14\,\Big[ N(N+2) + \bar{N}(\bar{N} + 2) -2N\bar{N}\Big]\,,
\ee
we get that in the $\rm{AdS_4}$ case with relations \p{ex}-\p{om-ads} the operator \p{ap-D-1}
takes the form:
\bea \lb{ap-D-2}
(\Pi)^2 &=& S(N,\bar{N}) L_0 + L^+ L + \left(\boldsymbol{\mu} +\frac{K^2}{4R}\right) L + \left(\boldsymbol{\mu} +\frac{(K-2)^2}{4R}\right) L^+
\\ [6pt]
&&{}+\frac{1}{R^2}\,A(\mathcal{M})+ \frac{1}{R^2} \,  (1 +\boldsymbol{\mu} R)
\bigl(\boldsymbol{\mu} R - \mathcal{M}_{m n} \mathcal{M}^{mn}\bigl)\,,
\nonumber
\eea
where $S(N,\bar{N})$ and $A(\mathcal{M})$ are defined in \p{A-C4PI} and \p{S-ap-MM}, respectively.
As a result, the fourth order Casimir operator \p{C4PI} is written as
\be \lb{C4PIf}
\mathcal{C}_4 = {}  R^2 \left[ \frac{1}{4} K(K-2) L_0 + L^+ L +  \left(\boldsymbol{\mu}
+\frac{K^2}{4R}\right) L + \left(\boldsymbol{\mu} +\frac{(K-2)^2}{4R}\right) L^+ \right]  +
\boldsymbol{\mu} R(1 + \boldsymbol{\mu} R)\,.
\ee
One can prove that the operator \p{C4PIf} is Hermitian. We see that the operator $\mathcal{C}_4$,
as well as the operator $\mathcal{C}_2$, is expressed through the operator constraints \p{L-0}-\p{L-p}.

\subsection{Irreducible representation}

Using the expressions \p{L0-C2} and \p{C4PIf},
we can determine the action of the $\mathfrak{so}(2,3)$ Casimir operators on
the infinite spin vectors \p{Psi} in $\rm{AdS_4}$ space satisfying the conditions \p{IrRep-1}.

From the first condition $L_0 {\Phi} = 0$ of the set  \p{IrRep-1}
we obtain that the infinite spin field \p{Psi} is the eigenvector of the second-order
Casimir operator
\be \lb{ev-QC}
\mathcal{C}_2 {\Phi} = c_2 {\Phi}\,
\ee
with the eigenvalue
\be \lb{ev-QC-2}
c_2= -2(1 + \boldsymbol{\mu} R)\,.
\ee

Besides, due to the first three conditions in \p{IrRep-1}, namely, the conditions
$L_0 {\Phi} = 0$, $L {\Phi} = 0$, $L^+ {\Phi} = 0$,
the same field ${\Phi}$ is also the eigenvector of the fourth order Casimir operator
\be \lb{ev-FC}
\mathcal{C}_4 {\Phi} = c_4 {\Phi}
\ee
with the eigenvalue
\be \lb{ev-FC-4}
c_4= \boldsymbol{\mu} R (1+ \boldsymbol{\mu} R)\,.
\ee

Thus, the field \p{Psi} under the conditions \p{IrRep-1} is the eigenvector of all Casimir operators
of the $\mathfrak{so}(2,3)$ algebra with the continuous eigenvalues \p{ev-QC-2} and \p{ev-FC-4}.
Therefore, the field $\Phi$ describing the infinite spin field dynamics in $\rm{AdS_4}$
space corresponds to the continuous irreducible representation of the $\mathrm{SO}(2,3)$ group.
The last condition in  \p{IrRep-1} removes the degeneracy of the spectrum associated with the use of
additional spinor operators.

Relations \p{ev-QC-2} and \p{ev-FC-4} show that
the eigenvalues of the Casimir operators are related to each other by the relation
\be \lb{c4-c2}
c_4= \frac{c_2}{2}\left(\frac{c_2}{2} +1 \right)
\ee
and are characterized by one dimensionless continuous real parameter $\boldsymbol{\mu} R$. This means
that of the two Casimir operators, only one is independent
\be
\lb{one casimir}
\mathcal{C}_4 = \frac{1}{2}\,\mathcal{C}_2\left(\frac{1}{2}\,\mathcal{C}_2 + 1\right) .
\ee
Such representations, where of all Casimir operators only one is independent,
are called the most degenerate representations in the mathematical literature. Therefore, the
Fock space vector describing the Lagrangian dynamics of the infinite spin field in $\rm{AdS_4}$ space
corresponds to the most degenerate continuous unitary representation of the $\rm{AdS_4}$ symmetry group.

In the paper \cite{Met2}, using the light-cone formalism, Metsaev proposed a classification of classically
unitary infinite spin fields in $\rm{AdS}$ space. The classification is based on relations (3.19) and  (3.20)-(3.24)
from \cite{Met2} for complex coefficients $p$ and $q$
in terms of which the eigenvalues of the second- and fourth-order Casimir operators in $\rm{AdS}_4$ space take the form \cite{Met2}:
\be\lb{Casimir-Metsaev}
c_2=p^2+q^2-\frac52\,,\qquad
c_4=\left(p^2-\frac14 \right)\left(q^2-\frac14 \right),
\ee
which are symmetric with respect to replacements $p\leftrightarrow q$.
We see that our obtained results \p{ev-QC-2}, \p{ev-FC-4} correspond to
\be\lb{our-Metsaev}
p=\varepsilon_p\left(\frac12+i \sqrt{\boldsymbol{\mu}R}\right),\qquad
q=\varepsilon_q\left(\frac12-i \sqrt{\boldsymbol{\mu}R}\right),
\ee
where $\varepsilon_p=\pm 1$ and $\varepsilon_q=\pm 1$ independently of each other.
In the case $\boldsymbol{\mu}>0$, the values \p{our-Metsaev} obtained are included in the set of solutions
$\mathbf{ii}$ and $\mathbf{iii}$ from \cite{Met2}, whereas  \p{our-Metsaev}
are the solutions from the set $\mathbf{vi}$ in the case $\boldsymbol{\mu}<0$, that is, $p$ and $q$ become real.
However, the cases $\mathbf{ii}$, $\mathbf{iii}$, $\mathbf{vi}$ as well as the other cases in \cite{Met2}
mean that two Casimir operators $\mathcal{C}_2$ and $\mathcal{C}_4$ are independent while
in our case only one of them is independent. As we have already pointed out, our case corresponds to the most
degenerate representation. Perhaps this degeneration is due to a special form of operator constraints
(\ref{L-0}), (\ref{L-n}), (\ref{L-p}).

\setcounter{equation}{0}

\section{Summary}

In this paper we have studied the algebraic aspects of the
Lagrangian formulation for the infinite spin field model proposed in \cite{BFIK-24_1}.
The main elements of our analysis are the fields
${\Phi}$
(\ref{Psi})  and the operator constraints (\ref{L-0})-(\ref{U}) acting on these fields and forming a closed algebra
in terms of commutators, which ensures gauge invariance of the
Lagrangian theory. The operator constraints in the case under consideration includes
the radius $R$ of $\rm{AdS}_4$ space and the arbitrary real massive parameter $\boldsymbol{\mu}$
associated with the continuous representation. It was proved that the above operator constraints are the
invariants of the $\mathrm{SO}(2,3)$ group and the Casimir operators of the  $\mathfrak{so}(2,3)$ algebra in the space
of vectors (\ref{Psi})
are completely expressed through these constraints in the form (\ref{L0-C2}) and (\ref{C4PIf}).

It turned out that the Casimir operators are not independent but are related to each other by
(\ref{one casimir}). Since there is only one independent Casimir operator in the given representation, we
conclude that the representation under consideration is most degenerate. One can show that the
constructed representations correspond to the cases $\mathbf{ii}$, $\mathbf{iii}$ (when $\boldsymbol{\mu}>0$)
and $\mathbf{vi}$  (when $\boldsymbol{\mu}<0$) in the Metsaev classification of
the unitary continuous spin irreducible representations in $\rm{AdS}$ space \cite{Met2}.

\section*{Acknowledgments}

 Authors are grateful to R.R.\,Metsaev for stimulating discussions and comments. The authors are also thankful to M.A.\,Vasiliev, Yu.M.\,Zinoviev and V.A.\,Krykhtin for useful comments. M.A.P. is grateful to A.A.\,Golubtsova for helpful discussions. The work of M.A.P. was supported by the JINR AYSS Foundation, Project No 24-301-03.

\end{document}